\documentclass[12pt,amsmath,amssymb,nofootinbib,superscriptaddress]{revtex4-1}
\usepackage{graphicx} 
\usepackage{subfigure} 
\usepackage{amsmath} %
\usepackage{pstricks}
\usepackage{color} 
\usepackage{amssymb}
\usepackage{slashed}
\usepackage{graphicx,amsfonts,layout,appendix,subfigure}
\input{epsf} 
\usepackage{verbatim} 
\usepackage{upgreek}
\usepackage{float} 

\def\beq{\begin{equation}}
\def\eeq{\end{equation}}
\def\bea{\begin{eqnarray}}
\def\eea{\end{eqnarray}}
\def\beqn{\begin{eqnarray}} 
\def\eeqn{\end{eqnarray}}

\def\nn{\nonumber}

\def\ln#1{\mathrm{ln}\left(#1\right)}

\newcommand\alphas{\alpha_{\mathrm{S}}}
\newcommand\as{a_{\mathrm{S}}}
\newcommand\al{a}

\newcommand\aso{a_{\mathrm{S}, 0}}

\newcommand{\matriz}[1]{\mathbf{#1}}

\def\beq{\begin{equation}} \def\eeq{\end{equation}}
\def\beqn{\begin{eqnarray}} \def\eeqn{\end{eqnarray}}
 
\def\nn{\nonumber}

\begin{document} 

\newcommand\sss{\scriptscriptstyle}
\newcommand\IDC{\textbf{\textit{{\rm Id}}}_C}
\newcommand\SUNT{\textit{\textbf{T}}}
\newcommand\pslashed{\slashed{p}}
\newcommand\DST{D_{\rm ST}}
\newcommand\DDirac{D_{\rm Dirac}}
\newcommand\factor{\rm Factor}
\newcommand\Spmatrix{\textit{\textbf{Sp}}}
\newcommand\Spelement{{\rm Sp}}
\newcommand\Spfunction{{\rm Split}}
\newcommand\NC{N_C}
\newcommand\CG{c_{\Gamma}}
\newcommand\DR{D_{\rm R}}
\newcommand\aem{\alpha_{\rm em}} \newcommand\refq[1]{$^{[#1]}$}
\newcommand\avr[1]{\left\langle #1 \right\rangle}
\newcommand\lambdamsb{\Lambda_5^{\rm \sss \overline{MS}}}
\newcommand\MSB{{\rm \overline{MS}}} \newcommand\MS{{\rm MS}}
\newcommand\DIG{{\rm DIS}_\gamma} \newcommand\CA{C_{\sss A}}
\newcommand\DA{D_{\sss A}} \newcommand\CF{C_{\sss F}}
\newcommand\TF{T_{\sss F}} 
\newcommand\qeps{q^2_{\epsilon}} 

\begin{titlepage}
\renewcommand{\thefootnote}{\fnsymbol{footnote}}
\begin{flushright}
     \end{flushright}
\par \vspace{10mm}

\begin{center}
{\Large \bf
QED corrections to parton distributions and Altarelli-Parisi splitting functions in the polarized case}
\end{center}

\par \vspace{2mm}
\begin{center}
{\bf Daniel de Florian}~$^{(a)}$\footnote{{\tt deflo@unsam.edu.ar}}, {\bf Lucas Palma Conte}~$^{(a)}$\footnote{{\tt lpalmaconte@unsam.edu.ar}}

\vspace{5mm}

${}^{(a)}$
International Center for Advanced Studies (ICAS), ICIFI \& ECyT-UNSAM, 25 de Mayo y Francia,
(1650) San Mart\'\i n,  Pcia. Buenos Aires, Argentina

\vspace{5mm}

\end{center}

\par \vspace{2mm}
\begin{center} {\large \bf Abstract} \end{center}
\begin{quote}
\pretolerance 10000

We discuss the effect of QED corrections in the evolution of polarized parton distributions.
We solve the corresponding evolution equations exactly to  ${\cal O}(\alpha )$ and ${\cal O}(\as^2)$ in Mellin $N$-space, extending the available techniques for pure QCD evolution. To accomplish this, we introduce, for the first time, the Altarelli-Parisi polarized kernels at LO in QED.
Furthermore, we perform a phenomenological analysis of the QED effects on polarized parton distributions (pPDFs), proposing different scenarios for the polarized photon density. Finally, we quantify the impact of the corresponding QED contributions to the polarized structure function $g_1$. We show that the relative corrections to both the pPDFs and the $g_1$ structure function are approximately at the few percent level, which is the order of magnitude expected considering the value of $\alpha$.
\end{quote}
\begin{flushleft}

\end{flushleft}
\end{titlepage}

\setcounter{footnote}{0}


\setcounter{footnote}{0}
\renewcommand{\thefootnote}{\fnsymbol{footnote}}

\section{Introduction}
\label{sec:introduction}
During the last few decades, there has been a great deal of activity in the area of high-energy physics with polarized processes, both from the experimental and theoretical points of view. The interest in this area arises from the need to understand the way in which the proton spin emerges from the share of its constituents, an open question and a key area of research in particle physics.
The spin content of the proton can be encoded in terms of the polarized distributions of quarks and gluons (pPDFs) \cite{Leader:2013uga}, which can be probed experimentally in high-energy collision processes with polarized nucleons. Despite the significant progress, due to the lack of measurements of observables probing a large kinematical range, associated with the challenges involved in the production of polarized beams, the situation is still unclear, particularly concerning the polarized gluon distribution.

Since early measurements at the polarized high-energy colliders began, the fixed-target polarized lepton-nucleon deep inelastic scattering (DIS) experiments performed have shown that a relatively small amount of the proton spin is carried by the quark and antiquark spins \cite{Nocera:2012hx,
Blumlein:2010rn,
Leader:2010rb,
Hirai:2008aj,deFlorian:2008mr,deFlorian:2009vb}
. However, these types of measurements are poorly sensitive to gluons. Instead, the best $\Delta g$ probes have so far been provided by the polarized proton-proton collisions available at the Relativistic Heavy Ion Collider at Brookhaven National Laboratory \cite{Aschenauer:2015eha}, where various processes such as jet or hadron production at high $p_T$ transverse momentum receive substantial contributions from gluon-induced hard scattering. These measurements improved the description of the gluon spin distribution and allowed a better understanding of the proton spin structure. In particular, global analyses performed by the DSSV \cite{deFlorian:2014yva,DeFlorian:2019xxt} and  NNPDF groups \cite{Nocera:2014gqa}, conclude that the contribution of $\Delta g$ to the proton spin is not negligible, although providing constraints only for a reduced range of proton momentum fractions.

In this context, the future  Electron-Ion Collider (EIC), which will enable a much wider kinematic range and achieve unprecedented precision for polarized measurements \cite{Accardi:2012qut,AbdulKhalek:2021gbh}, is expected to provide new insights into the spin share of the proton in terms of its fundamental components. 
Considering the forthcoming measurements, it is essential to increase the precision of the theoretical calculations. 
Since QCD computations are reaching high levels of accuracy, even N3LO in some cases, other effects that were not previously taken into account, e.g. QED effects, start to play a more significant phenomenological role in the theoretical predictions.
Given that $\alpha \approx \alpha_s^2$, NLO QED corrections compete with those at NNLO in QCD. These corrections typically manifest at the few percent levels for many observables, ultimately becoming quantitatively significant for achieving an accurate description.

In addition to the need to advance in the perturbative computations, it is also important to reach the same level of accuracy on the non-perturbative side, i.e., on the unpolarized and polarized parton distribution functions (PDFs and pPDFs). As mentioned above, the study of PDFs is essential to understand the proton structure. 
Although PDFs cannot be derived from first principles and are instead fitted using experimental data, their evolution with respect to energy can be computed. This calculation requires the knowledge of the Altarelli-Parisi splitting functions (AP kernels)\cite{Altarelli:1977zs}.
The computation of the NNLO corrections to the unpolarized splitting functions performed in \cite{Moch:2001im, Moch:2004pa, Vogt:2004mw, Vogt:2005dw} and the development of modern parton
distribution analyses \cite{ NNPDF:2021njg, Hou:2019efy, Bailey:2020ooq, H1:2021xxi,PDF4LHCWorkingGroup:2022cjn}, allows achieving the required accuracy in QCD. Also, the unpolarized kernels at LO in QED were calculated in \cite{Roth:2004ti, Kripfganz:1988bd,Spiesberger:1994dm}, and at NLO in QED and mixed order QED$\otimes$QCD in \cite{deFlorian:2016gvk,deFlorian:2015ujt}.
Solutions to the evolution equations including corrections at LO in QED were presented in \cite{Roth:2004ti,Bertone:2013vaa,Sadykov:2014aua} and different fits of the parton distributions including these corrections were performed in \cite{Carrazza:2015dea,Martin:2004dh,Ball:2013hta}.

A key density that appears for the first time when QED corrections are taken into account is the photon distribution, i.e, the probability of finding a photon in the nucleon. Previous analyses of this density were performed in \cite{Harland-Lang:2016kog,Schmidt:2015zda,Martin:2004dh,Ball:2013hta}, either obtaining large uncertainties or relying on phenomenological models for the contribution to the photon PDF from the low-energy regions \cite{Manohar:2016nzj}.  
In this context, a new approach towards obtaining an accurate description of this distribution was presented in \cite{Manohar:2016nzj, Manohar:2017eqh}
, where it is shown that the photon density can be computed from the present knowledge of the unpolarized structure functions.

As stated before, the situation is quite different in the polarized case. While the QCD splitting functions are known at LO \cite{Altarelli:1977zs}, NLO\cite{Mertig:1995ny,Vogelsang:1995vh,Vogelsang:1996im} and NNLO \cite{Moch:2014sna,Vogt:2014pha,Blumlein:2021enk}, the global analysis is still restricted to NLO QCD precision. To improve the accuracy in the polarized case, it is necessary to reach NNLO accuracy in QCD and include the QED corrections, as well.

In this article, we present for the first time, the expressions for polarized splitting functions and the polarized structure function $g_1$ to order $\alpha$ in QED and study the phenomenological impact of the corresponding corrections after introducing different scenarios for the polarized photon density. Working on the Mellin space, with a procedure similar to the one developed in \cite{Vogt:2004ns,Mottaghizadeh:2017vef} we solved the evolution equations analytically. 

This paper is organized as follows. In section \ref{sec:notacion}, we present the evolution equations and establish the main notation for this article. In section \ref{sec:kernels}, we introduce, for the first time, the QED polarized kernels, and in section \ref{sec:solutions}, we perform the analytical solution of the evolution equations in  Mellin space. The phenomenological impact, including the introduction of different scenarios for the polarized photon distributions, is studied in  section \ref{sec:result}. Finally, in section \ref{sec:conclusions}, we present the conclusions of this work.

\vspace{-0.4cm}
\section{Splitting kernels and parton distribution basis}
\label{sec:notacion}

We start by writing down the general expression for the evolution of gluon, photon and quark distributions. In order to simplify the notation, we write the distributions and kernels in the {\it unpolarized} form. The polarized case is recovered by simply considering the corresponding polarized proton distribution
and splitting functions, i.e. basically adding $\Delta$ in all the expressions below.
\beqn
\frac{dg}{dt}&=& \sum_{j=1}^{n_f} P_{g q_j} \otimes  q_j+ \sum_{j=1}^{n_f} P_{g \bar{q}_j} \otimes \bar{q}_j +  P_{g g} \otimes g +  P_{g \gamma} \otimes \gamma \, , \nn
\\ \frac{d\gamma}{dt}&=& \sum_{j=1}^{n_f} P_{\gamma q_j} \otimes  q_j+ \sum_{j=1}^{n_f} P_{\gamma \bar{q}_j} \otimes \bar{q}_j + P_{\gamma g} \otimes g + P_{\gamma \gamma} \otimes \gamma \, ,\label{Eq:DGLAP}
\\\frac{dq_i}{dt}&=& \sum_{j=1}^{n_f} P_{q_i q_j} \otimes q_j + \sum_{j=1}^{n_f} P_{q_i \bar{q}_j} \otimes \bar{q}_j + P_{q_i g} \otimes g + P_{q_i \gamma} \otimes \gamma \, ,\nn
\eeqn
with $t=\ln{Q^2}$ ($Q$ being the factorization scale), $n_f$ is the number of active fermions and $P_{ij}$ the Altarelli-Parisi splitting functions in the space-like region. The analog evolution equations for antiquarks can be obtained by applying charge conjugation. Here, we use the conventional notation
\beq
(f\otimes g)(x) = \int_x^1 \, \frac{dy}{y} \, f\left(\frac{x}{y}\right)g(y) \, ,
\eeq
to indicate convolutions. We do not include the lepton distributions, since up to the order we reach here they basically factorize from the rest of the distributions\footnote{To $\cal{O}(\alpha)$ lepton distributions only couple, in a trivial way, to the photon density.}.
Along this work we will use the expressions for the splitting functions including QCD and QED perturbative corrections. In this sense, each kernel can be expressed as,
\vspace{-0.18cm}
\beqn
P_{ij} &=&\sum_{\{a,b\}} \as^a \, a^b \, P_{ij}^{(a,b)},
\label{expanP}
\eeqn
where the indices $(a,b)$ indicate the (QCD,QED) perturbative order of the calculation, with $\as \equiv \frac{\alphas}{4\pi}$ and $a \equiv \frac{\alpha}{4\pi}$. Due to the QED corrections, the kernels can depend on the electric charge of the initiating quarks (up or down type). In general, we have $P_{i,j}^{(n,1)} \sim e_q^2$ with at least some $i,j=q$.

The quark splitting functions are decomposed as
\beqn
P_{q_i \, q_k} &=& \delta_{ik} \, P^V_{qq} + P^S_{qq} \, ,\\
 P_{q_i \, \bar{q}_k} &=& \delta_{ik} \, P^V_{q\bar{q}} + P^S_{q\bar{q}} \, ,\\
  P_q^{\pm} &=& P^V_{qq} \pm P^V_{q\bar{q}} \, ,
  \label{Eq:Pv}
\eeqn
which act as a definition for $P^V_{q q}$ and $P^V_{q \bar{q}}$. In order to minimize the mixing between the different parton distributions in the evolution, it is convenient to introduce the following basis, differentiating the following singlet and the non-singlet pPDF combinations \cite{Roth:2004ti}:
\beqn
\label{eq:basis}
f^{NS}&=&\{u_v,d_v,s_v,c_v,b_v,
\Delta_{uc},\Delta_{ds},\Delta_{sb}\}, \nn \\
 f^{S}&=&\{\Delta_{UD}, \Sigma,g,\gamma\ \} ,
\eeqn
where
\beqn
q_{v_i} &=& q_i-\bar{q_i} \, , \\
\Delta_{uc} &=& u+\bar{u}-c-\bar{c} \, ,\nn \\ 
\Delta_{ds} &=& d+\bar{d}-s-\bar{s} \, ,\nn \\ 
\Delta_{sb} &=& s+\bar{s}-b-\bar{b} \, , \\ 
\Delta_{UD} &=& u+\bar{u}+c+\bar{c} -d-\bar{d} -s-\bar{s}-b-\bar{b} \, , \\
 \Sigma &=& \sum_{i=1}^{n_f} ( q_i+\bar{q}_i)  \, .
\eeqn
$\Delta_{UD} $ could also include the top quark distribution in case of a 6 flavour analysis (adding $\Delta_{ct}$ and $t_v$ to complete the basis). 

Taking into account that beyond NLO in QCD the singlet {\it non-diagonal terms} ($P^S_{q \bar{q}}$ and $P^S_{qq}$) differ \cite{Catani:2004nc}, it is useful to define 
\beqn
\Delta P^S & \equiv & P^S_{qq} -P^S_{q \bar{q}}  , \nn \\
 P^S & \equiv &  P^S_{qq} + P^S_{q \bar{q}}, 
 \label{Eq:Ps}
\eeqn
where we explicitly use that these contributions do not depend on the quark charge up to the order we reach, since they do not receive QED corrections to $\cal{O}(\alpha)$.
In the expansion in powers of the couplings, the flavor-diagonal (‘valence’) quantity $P_{qq}^V$ in Eq.(\ref{Eq:Pv}) begins at first order. On the other hand, $P_{q\bar{q}}^V$ and the flavor-independent (‘sea’) contributions $P^S_{qq}$ and $P^S_{q\bar{q}}$, and therefore the ‘pure-singlet’ term $P^S$ in Eq.(\ref{Eq:Ps}), are of order $\alpha_s^2$. The kernel $\Delta P^S $ in Eq.(\ref{Eq:Ps}) arises for the first time at the third order. In this work, we reach NLO accuracy in QCD and LO in QED, which implies that $\Delta P^S = 0$ for all our calculations.

The evolution equations Eq.(\ref{Eq:DGLAP}) for the parton distributions in the basis of Eq.(\ref{eq:basis}) read for the non-singlet sector,
\beqn
\frac{dq_{v_i}}{dt} &=&   P_{q_i}^-     \otimes q_{v_i}  +\sum_{j=1}^{n_f} \Delta P^S  \otimes  q_{v_j}    \, ,
\label{eq:evolucionqvSIMPLE}
\\ \frac{d \{ \Delta_{uc} , \Delta_{ct} \}}{dt} &=&  P_{u}^+ \otimes \{ \Delta_{uc} , \Delta_{ct}  \} \, ,
\label{eq:evolucionDupperSIMPLE}
\\ \frac{d \{ \Delta_{ds} ,\Delta_{sb}  \}}{dt} &=&  P_{d}^+ \otimes \{ \Delta_{ds} ,\Delta_{sb}  \} \, ,
\label{eq:evolucionDlowerSIMPLE}
\eeqn
where in the second and third lines we use the notation $P^+_{u(d)}$, differentiating the kernels in two groups, those corresponding to the {\it up} quark flavours  ($u, \, c, \, t $)  and those corresponding to the {\it down} quark flavours ($d, \, s, \, b$).
As noted above, the term $\Delta P^S$ is zero up to the order we work, therefore, in the non-singlet sector we have uncoupled equations for the distributions $q_{v_i}$ (Eq.(\ref{eq:evolucionqvSIMPLE})). For the singlet sector we have, 
\beqn
\frac{d  \Delta_{UD}  }{dt} &=&  \frac{P_{u}^+ + P_{d}^+}{2} \otimes \Delta_{UD} 
                                                  + \frac{P_{u}^+ - P_{d}^+}{2} \otimes \Sigma +  (n_u-n_d) P^S \, \otimes \Sigma\nn 
                                                  \\ &+& 2 (n_u P_{ug} -n_d P_{dg})  \otimes g + 2 (n_u P_{u\gamma} -n_d P_{d\gamma})  \otimes \gamma \, ,
                                                  \label{eq:evolucionDUDSIMPLE}
\\ \nn 
\frac{d  \Sigma  }{dt} &=&  \frac{P_{u}^+ + P_{d}^+}{2} \otimes \Sigma
                                                  + \frac{P_{u}^+ - P_{d}^+}{2} \otimes \Delta_{UD} +  n_f \, P^S \otimes \Sigma\nn 
                                                  \\ &+& 2 (n_u P_{ug} +n_d P_{dg})  \otimes g + 2 (n_u P_{u\gamma} +n_d P_{d\gamma})  \otimes \gamma \, .
                                                  \label{eq:evolucionSIGMASIMPLE}
\\
\frac{d  g  }{dt} &=&  P_{g q} \otimes \Sigma  +P_{g g} \otimes g \, .
                                                  \label{eq:evolucionGLUONSIMPLE}
                                                  \\
\frac{d  \gamma  }{dt} &=&\frac{P_{\gamma u}-P_{\gamma d}}{2} \otimes \Delta_{UD} +\frac{P_{\gamma u}+P_{\gamma d}}{2}\otimes \Sigma + P_{\gamma\gamma} \otimes \gamma  \, .
                                                  \label{eq:evolucionGAMMASIMPLE}
\eeqn
Notice that in the limit of an equal number of $u$ and $d$ quarks ($n_u=n_d$) and the same electric charges ($P_{ug}=P_{dg}, P_{u\gamma}=P_{d\gamma}$, $P_{u}^+=P_{d}^+$), $\Delta_{UD}$ decouples from the other distributions in the evolution, while the singlet evolution recovers the usual pure-QCD expression. 

\section{QCD-QED polarized splitting kernels}
\label{sec:kernels}
Since the calculation at LO in QED is analogous to that in QCD, except for the colour factors and electric charges, the LO Altarelli-Parisi kernels in QED can be derived from those in QCD by adjusting the colour factors as follows \cite{deFlorian:2016gvk, deFlorian:2015ujt} 
\begin{equation}
C_A \rightarrow 0, \;\;\;\;\;\;C_F \rightarrow e_q^2, \;\;\;\;\;\; T_R \rightarrow N_C \, e_q^2.\;\;\;\;\;\;
\label{eq:colorajust}
\end{equation}
And, if the Feynman diagram involves a loop of fermions, as in the case of $\Delta P_{gg}^{(1,0)}$, 
one needs to apply the replacement:
\begin{equation}
n_f \, T_R  \rightarrow \sum_{f} e_f^2,
\label{eq:colorajust2}
\end{equation}
where the sum runs over all the fermions involved in the process.
To set the correct normalization, we start by reminding the lowest order polarized splitting functions in QCD $\Delta P_{ij}^{(1,0)}$ \cite{Altarelli:1977zs}
\beqn
\nn \Delta P_{qq}^{(1,0)}(x) &=& 2 \, C_F \left[ \frac{1+x^2}{(1-x)_+}  +\frac{3}{2}  \delta (1-x) \right] \equiv 2 \,C_F\, \left[  \delta p_{qq}(x) + \frac{3}{2} \delta(1-x)\right] \, , 
\\ \nn \Delta P_{qg}^{(1,0)}(x) &=& T_R \left[   4x-2\right] \equiv T_R \, \delta  p_{qg}(x) \, , 
\\  \Delta P_{gq}^{(1,0)}(x) &=& C_F \left[   4-2x  \right] \equiv C_F \,  \delta p_{gq}(x) \, , \nn \\
\Delta P_{gg}^{(1,0)}(x) &=& 2 \, C_A \left[  \frac{1}{(1-x)_+}   -2x+1 \right] + \beta_0 \, \delta(1-x) \equiv 2 \, C_A \, \delta p_{gg}(x) +  \beta_0 \, \delta(1-x) \, ,\label{Eq:kernelsQCD}
\label{Eq:kernelsQCD}
\eeqn
with $\beta_0=\frac{11 N_C-4 n_f T_R}{3}$ and the usual plus distribution is defined as
\beq
\int_0^1 dx \, \frac{f(x)}{(1-x)_+} = \int_0^1 dx \, \frac{f(x)-f(1)}{1-x} \, ,
\eeq
for any regular test function $f$.
Therefore, setting the  factors as Eq.(\ref{eq:colorajust}), the lowest order splitting functions in QED $P_{ij}^{(0,1)}$ result in
\beqn
\Delta P_{qq}^{(0,1)}(x) &=& 2 \, e_q^2 \left[  \delta  p_{qq}(x) +\frac{3}{2} \delta (1-x) \right] \, , \nn \\
\Delta P_{q\gamma}^{(0,1)}(x) &=& N_C \, e_q^2  \, \delta  p_{qg}(x) \, ,  \nn \\
\Delta P_{\gamma q}^{(0,1)}(x) &=&  e_q^2 \, \delta  p_{gq}(x) \, , \label{Eq:kernelsQED} \\
\Delta P_{\gamma\gamma}^{(0,1)}(x) &=& -\frac{4}{3} \sum_{f} e_f^2 \, \delta (1-x)  \, ,
\nn
\eeqn
where there is an explicit dependence on the quark electromagnetic charge.  Furthermore, the sum over fermion charges in the $\Delta P^{(0,1)}_{\gamma\gamma}$ kernel corresponds to the definition 
\beq
\sum_{f}e_f^2 =N_C \sum_{q}^{n_f}e_q^2 \, + \, \sum_{l}^{n_L} e_l^2 \, ,
\eeq
with $n_f$ and $n_L$ the number of quark and lepton flavours, respectively.

The expressions for NLO QCD corrections to the splitting functions $\Delta P_{ij}^{(2,0)}$ can be found in \cite{Mertig:1995ny,Vogelsang:1995vh,Vogelsang:1996im}.
\section{Solving the evolution equations}
\label{sec:solutions}

To solve the evolution equations, we work in Mellin $N$-space.  We define the Mellin transformation, from Bjorken $x$-space to
complex  $N$-moment space, as
\beqn
a(N)=\int_0^1  dx \, x^{N-1} \, a(x),
\label{Eq:mellin}
\eeqn
and its inverse reads
\beqn
a(x)=\frac{1}{2\pi i}\int_{\textit{C}_N}dN \, x^{-N} \, a(N),
\label{eq:antitransofr}
\eeqn
here $\textit{C}_N$ denotes a suitable contour in the complex $N$ plane that has an imaginary part ranging from $-\infty$ to $\infty$ and that intersects the real axis to the right of the rightmost pole of $a(N)$. In practice, it is beneficial to choose the contour to be bent at an angle  $< \pi/2$ towards the negative real-$N$ axis \cite{Vogt:2004ns,Gluck:1989ze}. The integration in Eq.(\ref{eq:antitransofr}) can then be very efficiently performed numerically by choosing the values of $N$ as the supports for a Gaussian integration \cite{Vogt:2004ns}.

The advantage of working in this space is that the convolutions appearing in the evolution equations can be written simply as products,
\beq
(a\otimes b)(N)=a(N) \, b(N),
\eeq
which makes it easier to manipulate the expressions and solve the corresponding equations analytically at a given order. From now on, all kernels and pPDFs will be expressed in this space, but we will keep the nomenclature a bit loose to avoid complicating the notation. 

\subsection{Non-Singlet case}

We start with the solution for the non-singlet distributions (Eqs.(\ref{eq:evolucionqvSIMPLE}),(\ref{eq:evolucionDupperSIMPLE}),(\ref{eq:evolucionDlowerSIMPLE})).
If we express the kernels with the form of Eq.(\ref{expanP}), we have  a generic form for the evolution
\beqn
\frac{df(Q)}{dt}=(\as P^{(1,0)}+\as^2 P^{(2,0)}+\al P^{(0,1)} + {\mathcal O}(\as^3,\al^2,\as \al) ) f(Q_0),
\label{eq:evolucion}
\eeqn
where $f$ is any distribution of the non-singlet sector, and $P$ is the kernel corresponding to the evolution of that distribution. The solution is customarily obtained by posing an evolution operator that can be decomposed in the product of an LO QED operator and a corresponding LO and NLO one in QCD  as,
\begin{equation}
f(Q)=E^{QCD}\left(Q, Q_0\right)E^{QED}\left(Q, Q_0\right) f\left(Q_0\right)=E^{(2,0)}\left(Q, Q_0\right) E^{(1,0)}\left(Q, Q_0\right)  E^{(0,1)}\left(Q, Q_0\right) f\left(Q_0\right).
\label{solu}
\end{equation}
We will make use of the well-known RGE for the strong and electromagnetic coupling constants,
\beqn
\frac{d\as}{dt}&=&-(\as^2 \beta_0 + \as^3 \beta_1 +...), \nn \\
\frac{d\al}{dt}&=&-(\al^2 \beta'_0  +...),
\label{eq:copuling}
\eeqn
to express the Eq.(\ref{eq:evolucion}) in terms of $\al$ and $\as$. Inserting the expression in Eq.(\ref{solu})  into Eq.(\ref{eq:evolucion}) we can find separate solutions for QED and QCD. Keeping the accuracy only up to NLO in QCD and LO in QED, we get 
\beqn
\frac{d E^{QCD}\left(Q, Q_0\right)}{d a_S}  &=&-\frac{\as P^{(1,0)} +\as^2 P^{(2,0)}}{\as^2(\beta_0+\as \beta_1)} E^{QCD}\left(Q, Q_0\right)\nn\\
&\simeq& -\frac{1}{\as \beta_0} \left[ P^{(1,0)} +\as \left( P^{(2,0)}-\frac{\beta_1}{\beta_0} P^{(1,0)}\right)\right] E^{QCD}\left(Q, Q_0\right), \label{eq:solucionesseparadas}
 \\
\frac{d E^{QED}\left(Q, Q_0\right)}{d a} &=&-\frac{1}{\al \beta'_0}  P^{(0,1)} E^{QED}\left(Q, Q_0\right).\nn
\eeqn

Let's start with the solution for the QCD sector, if we remain at LO accuracy we have
\beqn
\frac{dE^{(1,0)}(Q,Q_0)}{d\as} = -\frac{1}{\as \beta_0}  P^{(1,0)} \, E^{(1,0)}(Q,Q_0),
\eeqn
which, with the constraint $E^{LO}(Q_0,Q_0)=1$, we obtain the well known exponential solution
\beqn
E^{(1,0)}(Q,Q_0) = \left( \frac{\as}{\aso} \right)^{\frac{-P^{(1,0)}}{\beta_0}} ,
\eeqn  
where we define $\as \equiv \as(Q)$ and  $\aso\equiv \as(Q_0)$. For the NLO component one gets 
\beqn
\frac{dE^{(2,0)}(Q,Q_0)}{d\as} = -\frac{1}{\beta_0} \left( P^{(2,0)}-\frac{\beta_1}{\beta_0} P^{(1,0)}\right) \, E^{(2,0)}(Q,Q_0),
\eeqn
with the solution valid up to NLO accuracy as
\beqn
E^{(2,0)}(Q,Q_0) &=&\exp\left[ -\frac{\as-\aso}{\beta_0} \left( P^{(2,0)}-\frac{\beta_1}{\beta_0} P^{(1,0)}\right) \right] \nn\\
&=&
1 -\frac{\as-\aso}{\beta_0} \left( P^{(2,0)}-\frac{\beta_1}{\beta_0} P^{(1,0)}\right) +{\cal O}(\as^2).
\label {eq:nlo}
\eeqn
By combining the LO and NLO operators, we arrive at
\beqn
E^{QCD}\left(Q, Q_0\right)=\left( \frac{\as}{\aso} \right)^{\frac{-P^{(1,0)}}{\beta_0}}\left(
1 -\frac{\as-\aso}{\beta_0} \left( P^{(2,0)}-\frac{\beta_1}{\beta_0} P^{(1,0)}\right)\right)+{\cal O}(\as^2).
\eeqn
The same procedure can be applied to the QED sector as well. Keeping up to LO, the resulting QED evolution operator is given by
\beqn
E^{QED}\left(Q, Q_0\right)=E^{(0,1)}\left(Q, Q_0\right)=\left(\frac{\al}{\al_0}\right)^{-\frac{P^{(0,1)}}{\beta'}},
\label{EVOLQEDnonsinglet}
\eeqn
where we define $\al \equiv \al(Q)$ and $\al_0 \equiv \al(Q_0)$.
As long as all flavours are active, all combinations evolve as non-singlets. But if one quark flavour is not active, the corresponding $\Delta_{ij}$ combination containing that quark term can be expressed as a combination of singlet and non-singlet terms.
(e.g. at $Q=1$ GeV there is no charm, so we express $\Delta_{uc}=\frac{1}{2}(\Sigma+\Delta_{UD})$).

\subsection{Singlet case}
For the singlet sector, we can express the evolution equations (Eqs.(\ref{eq:evolucionDUDSIMPLE}), (\ref{eq:evolucionSIGMASIMPLE}), (\ref{eq:evolucionGLUONSIMPLE}) and (\ref{eq:evolucionGAMMASIMPLE})) in a matrix form,
\beqn
\frac{d \bar{f^S}}{d t}= \matriz{P} . \bar{f^S},
\label{eq:singlete}
\eeqn
where $\bar{f^S}$ is the vector of singlet combinations of flavors Eq.(\ref{eq:basis}) and $\matriz{P}$ is the matrix of kernels,

\begin{equation}
\matriz{P}=\left(\begin{array}{cccc}
\frac{P_u^++P_d^+}{2} & \frac{P_u^+-P_d^+}{2}+(n_u-n_d)P^S & 2(n_u P_{ug}-n_d P_{dg}) & 2(n_u P_{u\gamma}-n_d P_{d\gamma}) \\
\frac{P_u^+-P_d^+}{2}  & \frac{P_u^++P_d^+}{2}+n_f P^S  & 2(n_u P_{ug}+n_d P_{dg}) & 2(n_u P_{u\gamma}+n_d P_{d\gamma})  \\
0 & P_{gq} & P_{gg} & 0 \\
\frac{P_{\gamma u}-P_{\gamma d}}{2} &\frac{P_{\gamma u}+P_{\gamma d}}{2} & 0 & P_{\gamma\gamma}
\end{array}\right),
\end{equation}
the kernel matrix can be expressed in a perturbative form,
\beqn 
\matriz{P}=\as \matriz{P}^{(1,0)}+\as^2 \matriz{P}^{(2,0)}+\al \matriz{P}^{(0,1)} +\cdots . 
\label{eq:Pmatriz}
\eeqn
For QCD matrices, with no photon content as noticed in the last column/row, we have,
\beqn
\matriz{P}^{(1,0)}=\left(\begin{array}{cccc}
P_{q q}^{(1,0)} & 0 &2 \, n_{ud}P_{q g}^{(1,0)} & 0 \\
0 & P_{q q}^{(1,0)} & 2\, n_f P_{q g}^{(1,0)} & 0 \\
0 & P_{g q}^{(1,0)} & P_{g g}^{(1,0)} & 0 \\
0 & 0 & 0 & 0
\end{array}\right), \, \matriz{P}^{(2,0)}=\left(\begin{array}{cccc}
P^{+^{(2,0)}} & n_{ud} P^{S^{(2,0)}} &2 \, n_{ud}P_{q g}^{(2,0)} & 0 \\
0 & P_{q q}^{(2,0)} &2 \, n_f P_{q g}^{(2,0)} & 0 \\
0 & P_{g q}^{(2,0)} & P_{g g}^{(2,0)} & 0 \\
0 & 0 & 0 & 0
\end{array}\right),\eeqn
where $n_{ud}\equiv n_u-n_d$ and we use that to this order, we can express:
\beqn
P_{ug}^{(i,0)}&=&P_{dg}^{(i,0)}= P_{qg}^{(i,0)}, \nn \\
P_{gu}^{(i,0)}&=&P_{gd}^{(i,0)}= P_{gq}^{(i,0)}, \nn \\
P^{+^{(1,0)}}_d&=&P^{+^{(1,0)}}_u= P^{(1,0)}_{qq}, \nn\\
P^{+^{(2,0)}}_u&=&P^{+^{(2,0)}}_d\equiv P^{+^{(2,0)}},\nn\\
P_{qq}^{(2,0)}&\equiv& P^{+(2,0)} + n_f P^{S(2,0)},\nn
\eeqn
where in the first two lines $i=1,2$. The expression of the kernels $P^{(1,0)}_{qq}$, $P^{(1,0)}_{qg}$, $P^{(1,0)}_{gq}$, and $P^{(1,0)}_{gg}$ are obtained by performing the corresponding transform to Mellin $N$-space on Eq.(\ref{Eq:kernelsQCD}). In the third line, we use that $P_{q\bar{q}}^{V(1,0)}=0$, and to construct the LO matrix, we employ that $P^{S(1,0)}=0$. The NLO polarized kernels  $P_{qg}^{(2,0)}$, $P_{gq}^{(2,0)}$, $P^{(2,0)}_{gg}$, $P^{S(2,0)}$ and $P^{+^{(2,0)}}$ can be found in \cite{Mertig:1995ny,Vogelsang:1995vh,Vogelsang:1996im}. For the QED sector, with no gluon content in the third column/row, we have, 
\beqn
\matriz{P}^{(0,1)}=\left(\begin{array}{llll}
\eta ^+ P_{ff}^{(0,1)}& \eta ^- P_{ff}^{(0,1)} & 0 & \delta^2 P_{f\gamma}^{(0,1)}\\
\eta ^- P_{ff}^{(0,1)} & \eta ^+P_{ff}^{(0,1)} & 0 & e^2_\Sigma P_{f\gamma}^{(0,1)}\\
0 & 0 & 0 & 0\\
\eta ^- P_{\gamma f}^{(0,1)}& \eta ^+ P_{\gamma f}^{(0,1)} &0 & P_{\gamma\gamma}^{(0,1)}
\end{array}\right),
\label{Eq:Pqed}
\eeqn
where we use that to this perturbative order, we can express
\beqn
 P_{u(d)}^{\pm{(0,1)}}&=&e_{u(d)}^{2}P_{ff}^{(0,1)},\\
 P_{u(d)\gamma}^{(0,1)}&=&e_{u(d)}^2 P_{f\gamma}^{(0,1)},\\
 P_{\gamma u(d)}^{(0,1)}&=&e_{u(d)}^2 P_{\gamma f}^{(0,1)},
 \eeqn
with the definitions of $P_{ff}^{(0,1)}$,$P_{f\gamma}^{(0,1)}$,$P_{\gamma f}^{(0,1)}$ deduced from Eq.(\ref{Eq:kernelsQED}) by extracting the electric charge $e_q^2$ of the corresponding kernels and taking the corresponding transform to Mellin $N$-space. In the first line, we use that $P_{u(d)}^{\pm(0,1)}=\Delta P_{qq}^{(0,1)}$ from Eq.(\ref{Eq:kernelsQED}) with $e_q^2=e_u^2\,(e_d^2$) , which arises from the fact that $P^{V(0,1)}_{q\bar{q}}=0$ in Eq.(\ref{Eq:Pv}). 
In addition, we define
\beqn
\eta^{\pm}&=&\frac{1}{2}\, (e_u^2 \pm e_d^2),  \\
e^2_\Sigma&=&2\, (n_u \, e_u^2+n_d\, e_d^2),\\ 
\delta^2&=&2\, (n_u \, e_u^2-n_d\, e_d^2).
 \eeqn
One non-trivial problem arises from the fact that the evolution matrix in Eq.(\ref{eq:Pmatriz}) has non-commutative properties at different orders. Even the LO QCD matrix does not commute with the QED one. This makes more difficult to isolate the QED and QCD solutions than in the non-singlet case. However, if mixed-order terms ($\mathcal{O}(\al\,\as)$) are thrown away, which is correct at the order we are working, we can separate the  QED and QCD equations. A more detailed discussion is presented in section \ref{qedyqcd}. In summary, we obtain matrix equations analogous to Eq.(\ref{eq:solucionesseparadas}) that, in the case of QCD, correspond to
\beqn
\frac{d \matriz{E}^{QCD}\left(Q, Q_0\right)}{d a_S} \simeq -\frac{1}{\as \beta_0} \left[\matriz{P}^{(1,0)} +\as \left( \matriz{P}^{(2,0)}-\frac{\beta_1}{\beta_0} \matriz{P}^{(1,0)}\right)\right]\matriz{E}^{QCD}\left(Q, Q_0\right).
\label{eq:ecuacionSingleteQCD}
\eeqn
where the matrices $\matriz{P}^{(1,0)}$ and $\matriz{P}^{(2,0)}$ do not commute, so the equation cannot be solved in a closed exponential form. The usual way to solve the equation is to combine the exact solution at LO with a power expansion in the coupling for the rest \cite{Vogt:2004ns,Mottaghizadeh:2017vef}. The LO solution can be expressed as a closed exponential form,
\beqn
\matriz{E}_{LO}^{QCD}\left(Q, Q_0\right)=\left(\frac{\as}{\aso}\right)^{-\frac{\matriz{P^{(1,0)}}}{\beta_0}} .
\eeqn
Note that in the exponent we have the matrix $\matriz{P}^{(1,0)}$, to be able to operate in the following steps it is convenient to rewrite this expression by applying the eigenvalue decomposition of the LO splitting function matrix,
\beqn
\matriz{P}^{(1,0)}= \lambda_1 \, \matriz{e}_1+\lambda_2 \, \matriz{e}_2+\lambda_3 \,\matriz{e}_3+\lambda_4\, \matriz{e}_4,
\eeqn
where $\lambda_i$ are given by
\beqn
\lambda_{1,2}&=&\frac{1}{2}\left(P_{qq}^{(1,0)}+P_{gg}^{(1,0)}\pm \sqrt{(P_{qq}^{(1,0)}-P_{gg}^{(1,0)})^2+8\,n_f\,P_{qg}^{(1,0)}P_{gq}^{(1,0)}} \, \right),  \\
\lambda_3&=&P_{qq}^{(1,0)}, \hspace{1cm} \lambda_4=0, \nn
\eeqn
and the matrices $\matriz{e}_i$  are the projectors to the subspace corresponding to each eigenvector,

\beqn
\matriz{e}_1&=&\left(\begin{array}{cccc}
0 & \frac{n_{ud}\,(\lambda_2-P_{qq}^{(1,0)})}{n_f\,(\lambda_1-\lambda_2)} & \frac{n_{ud}\,(\lambda_2-P_{qq}^{(1,0)}) (P_{qq}^{(1,0)}-\lambda_1)}{n_f\,P_{gq}^{(1,0)}(\lambda_1-\lambda_2)} & 0 \\
0 & \frac{\lambda_2-P_{qq}^{(1,0)}}{\lambda_2-\lambda_1} &\frac{(\lambda_2-P_{qq}^{(1,0)}) (\lambda_1-P_{qq}^{(1,0)})}{P_{gq}^{(1,0)}(\lambda_2-\lambda_1)} & 0 \\
0 & \frac{P_{g q}^{(1,0)} }{\lambda_1-\lambda_2}& \frac{P_{qq}^{(1,0)}-\lambda_1}{\lambda_2-\lambda_1} & 0 \\ \nn
0 & 0 & 0 & 0
\end{array}\right),
\\ 
\matriz{e}_2&=&\left(\begin{array}{cccc}
0 & \frac{n_{ud}\,(\lambda_1-P_{qq}^{(1,0)})}{n_f\,(\lambda_1-\lambda_2)} & \frac{n_{ud}\,(\lambda_2-P_{qq}^{(1,0)}) (P_{qq}^{(1,0)}-\lambda_1)}{n_f\,P_{gq}^{(1,0)}(\lambda_2-\lambda_1)} & 0 \\
0 & \frac{\lambda_1-P_{qq}^{(1,0)}}{\lambda_1-\lambda_2} &\frac{(\lambda_2-P_{qq}^{(1,0)}) (\lambda_1-P_{qq}^{(1,0)})}{P_{gq}^{(1,0)}(\lambda_1-\lambda_2)} & 0 \\
0 & \frac{P_{g q}^{(1,0)} }{\lambda_2-\lambda_1}& \frac{P_{qq}^{(1,0)}-\lambda_2}{\lambda_1-\lambda_2} & 0 \\
0 & 0 & 0 & 0
\end{array}\right), 
\\
\matriz{e}_3&=&\left(\begin{array}{cccc}
1 & -\frac{n_{ud}}{n_f}&0& 0 \\
0 & 0&0 & 0 \\
0 & 0& 0 & 0 \\
0 & 0 & 0 & 0
\end{array}\right), \hspace{1cm} \matriz{e}_4=\left(\begin{array}{cccc}
0& 0 & 0 & 0 \\
0 & 0 &0 & 0 \\
0 & 0& 0 & 0 \\
0 & 0 & 0 & 1
\end{array}\right), \nn
\eeqn
with a bit of algebra, we arrive at the following expression for the evolution operator at LO,
\beqn
\matriz{E}_{LO}^{QCD}\left(Q, Q_0\right)=\matriz{e}_1\left(\frac{\as}{\aso}\right)^{-\frac{\lambda_1}{\beta_0}}+\matriz{e}_2\left(\frac{\as}{\aso}\right)^{-\frac{\lambda_2}{\beta_0}}+\matriz{e}_3\left(\frac{\as}{\aso}\right)^{-\frac{\lambda_3}{\beta_0}}+\matriz{e}_4\left(\frac{\as}{\aso}\right)^{-\frac{\lambda_4}{\beta_0}}.
\eeqn
As mentioned, we can propose a general solution of Eq.(\ref{eq:ecuacionSingleteQCD}) employing a series expansion around the  LO result, reading \cite{Vogt:2004ns}
\beqn
\matriz{E}^{QCD}\left(Q, Q_0\right)=\matriz{U}(\as) \, \matriz{E}_{LO}^{QCD}\left(Q, Q_0\right) \,\matriz{U}^{-1}(\aso),\hspace{1.5mm}\textup{ with}\hspace{1mm} \matriz{U}(\as)=1+\sum_{k=1}^\infty \as^k \, \matriz{U}_k.
\label{eq:fullsolu}
\eeqn 
Keeping terms up to NLO we have the truncated calculation
\beqn
\matriz{E}^{QCD} \simeq \matriz{L}+\as \matriz{U}_1\matriz{L}-\aso \matriz{L}\matriz{U}_1.
\label{Eq:trunsolu}
\eeqn 
If we replace this solution in Eq.(\ref{eq:ecuacionSingleteQCD}) we can obtain the following relation for the matrix $\matriz{U}_1$, 
\beqn
[\matriz{U}_1,\matriz{P}^{(1,0)}]=\matriz{R}+\matriz{U}_1 \textup{, with}\hspace{3mm} \matriz{R}=\left( \matriz{P}^{(2,0)}-\frac{\beta_1}{\beta_0} \matriz{P}^{(1,0)}\right).
\label{eq:Ucon}
\eeqn 

By decomposing the matrix $\matriz{U}_1$ into the eigenvalue subspaces of $\matriz{P}^{(1,0)}$ and using the relation Eq.(\ref{eq:Ucon}) we can find an expression for $\matriz{U}_1$ in terms of the matrix $\matriz{R}$ and the projectors $\matriz{e}_i$. Then from Eq.(\ref{Eq:trunsolu}) we obtain the evolution operator,
\beqn
&\matriz{E}_{QCD}\left(Q, Q_0\right)=\left\{( \frac { \as} { \aso } ) ^ { -\frac { \lambda _ { 1 } } { \beta _ { 0 }  } } \left(\matriz{e}_1-\frac{\left(\aso-\as\right)}{\beta_{0}} \matriz{e}_1\matriz{R}\matriz{e}_1-\frac{1}{\beta_{0}+\lambda_{2}-\lambda_{1}}\left(\as\left(\frac{\as}{\aso}\right)^{\left(\lambda_{2}-\lambda_{1}\right) / \beta_{0}}-\aso\right) \matriz{e}_1\matriz{R}\matriz{e}_2\right.\right.\nn \\
&\left.-\frac{1}{\beta_{0}+\lambda_{3}-\lambda_{1}}\left(\as\left(\frac{\as}{\aso}\right)^{\left(\lambda_{3}-\lambda_{1}\right) / \beta_{0}}-\aso\right) \matriz{e}_1\matriz{R}\matriz{e}_3-\frac{1}{\beta_{0}+\lambda_{4}-\lambda_{1}}\left(\as\left(\frac{\as}{\aso}\right)^{\left(\lambda_{4}-\lambda_{1}\right) / \beta_{0}}-\aso\right) \matriz{e}_1\matriz{R}\matriz{e}_4\right)\nn\\
&+\left(\frac{\as}{\aso}\right)^{-\frac{\lambda_{2}}{\beta_{0}}}\left(\matriz{e}_2-\frac{\left(\aso-\as\right)}{\beta_{0}} \matriz{e}_2\matriz{R}\matriz{e}_2-\frac{1}{\beta_{0}+\lambda_{1}-\lambda_{2}}\left(\as\left(\frac{\as}{\aso}\right)^{\left(\lambda_{1}-\lambda_{2}\right) / \beta_{0}}-\aso\right) \matriz{e}_2\matriz{R}\matriz{e}_1\right.\nn\\
&\left.-\frac{1}{\beta_{0}+\lambda_{3}-\lambda_{2}}\left(\as\left(\frac{\as}{\aso}\right)^{\left(\lambda_{3}-\lambda_{2}\right) / \beta_{0}}-\aso\right) \matriz{e}_2\matriz{R}\matriz{e}_3-\frac{1}{\beta_{0}+\lambda_{4}-\lambda_{2}}\left(\as\left(\frac{\as}{\aso}\right)^{\left(\lambda_{4}-\lambda_{2}\right) / \beta_{0}}-\aso\right) \matriz{e}_2\matriz{R}\matriz{e}_4\right)\nn\\
&+\left(\frac{\as}{\aso}\right)^{-\frac{\lambda_{3}}{\beta_{0}}}\left(\matriz{e}_3-\frac{\left(\aso-\as\right)}{\beta_{0}} \matriz{e}_3\matriz{R}\matriz{e}_3-\frac{1}{\beta_{0}+\lambda_{1}-\lambda_{3}}\left(\as\left(\frac{\as}{\aso}\right)^{\left(\lambda_{1}-\lambda_{3}\right) / \beta_{0}}-\aso\right) \matriz{e}_3\matriz{R}\matriz{e}_1\right.\nn\\
&\left.-\frac{1}{\beta_{0}+\lambda_{2}-\lambda_{3}}\left(\as\left(\frac{\as}{\aso}\right)^{\left(\lambda_{2}-\lambda_{3}\right) / \beta_{0}}-\aso\right) \matriz{e}_3\matriz{R}\matriz{e}_2-\frac{1}{\beta_{0}+\lambda_{4}-\lambda_{3}}\left(\as\left(\frac{\as}{\aso}\right)^{\left(\lambda_{4}-\lambda_{3}\right) / \beta_{0}}-\aso\right) \matriz{e}_3\matriz{R}\matriz{e}_4\right)\nn\\
&+\left(\frac{\as}{\aso}\right)^{-\frac{\lambda_{4}}{\beta_{0} }}\left(\matriz{e}_4-\frac{\left(\aso-\as\right)}{\beta_{0}} \matriz{e}_4\matriz{R}\matriz{e}_4-\frac{1}{\beta_{0}+\lambda_{1}-\lambda_{4}}\left(\as\left(\frac{\as}{\aso}\right)^{\left(\lambda_{1}-\lambda_{4}\right) / \beta_{0}}-\aso\right) \matriz{e}_4\matriz{R}\matriz{e}_1\right.\nn\\
&\left.\left.-\frac{1}{\beta_{0}+\lambda_{2}-\lambda_{4}}\left(\as\left(\frac{\as}{\aso}\right)^{\left(\lambda_{2}-\lambda_{4}\right) / \beta_{0}}-\aso\right) \matriz{e}_4\matriz{R}\matriz{e}_2-\frac{1}{\beta_{0}+\lambda_{3}-\lambda_{4}}\left(\as\left(\frac{\as}{\aso}\right)^{\left(\lambda_{3}-\lambda_{4}\right) / \beta_{0}}-\aso\right) \matriz{e}_4\matriz{R}\matriz{e}_3\right)\right\}. \nn
\eeqn
Since in the QED sector we restrict our analysis to LO, the evolution equation becomes simpler and can be expressed as follows
\beqn
\frac{d \matriz{E}^{QED}\left(Q, Q_0\right)}{d a} &=&-\frac{1}{\al \beta'_0}  \matriz{P}^{(0,1)} \matriz{E}^{QED}\left(Q, Q_0\right).
\label{eq:ecuacionSingleteQED}
\eeqn
Again, the solution has an exponential form and we can express it in terms of the eigenvalues of the matrix $\matriz{P}^{(0,1)}$  ($\lambda'_i$) and the projectors to the corresponding eigenvector subspace $\matriz{e}'_1$, $\matriz{e}'_2$, $\matriz{e}'_3$ and $\matriz{e}'_4$,
\beqn
\matriz{E}_{LO}^{QED}\left(Q, Q_0\right)=\left(\frac{\al}{\al_0}\right)^{-\frac{\matriz{P}^{(0,1)}}{\beta'_0}}=\matriz{e}'_1.\left(\frac{\al}{\al_0}\right)^{-\frac{\lambda_1'}{\beta'_0}}+\matriz{e}'_2\left(\frac{\al}{\al_0}\right)^{-\frac{\lambda_2'}{\beta'_0}}+\matriz{e}'_3\left(\frac{\al}{\al_0}\right)^{-\frac{\lambda_3'}{\beta'_0}}+\matriz{e}'_4\left(\frac{\al}{\al_0}\right)^{-\frac{\lambda_4'}{\beta'_0}},
\eeqn
with, 
\beqn
\matriz{e}'_1&=&\frac{-(\matriz{P}^{(0,1)})^2+\matriz{P}^{(0,1)}(\lambda'_2+\lambda'_3)-\lambda'_2\lambda'_3 \, \matriz{Id}_{4x4}+\matriz{e}'_4\, \lambda'_2\lambda'_3}{(\lambda'_1-\lambda'_3)(\lambda'_1-\lambda'_2)},\nn\\
\matriz{e}'_2&=&\frac{-(\matriz{P}^{(0,1)})^2+\matriz{P}^{(0,1)}(\lambda'_1+\lambda'_3)-\lambda'_1\lambda'_3 \, \matriz{Id}_{4x4}+\matriz{e}'_4\, \lambda'_1\lambda'_3}{(\lambda'_2-\lambda'_1)(\lambda'_2-\lambda'_3)},\\\nn
\matriz{e}'_3&=&\frac{-(\matriz{P}^{(0,1)})^2+\matriz{P}^{(0,1)}(\lambda'_1+\lambda'_2)-\lambda'_1\lambda'_2 \, \matriz{Id}_{4x4}+\matriz{e}'_4\, \lambda'_1\lambda'_2}{(\lambda'_2-\lambda'_3)(\lambda'_3-\lambda'_1)},\\\nn
\matriz{e}'_4&=&\left(\begin{array}{cccc}
0& 0 & 0 & 0 \\
0 & 0 &0 & 0 \\
0 & 0& 1 & 0 \\
0 & 0 & 0 & 0
\end{array}\right),\nn
\eeqn
where the eigenvalues are $\lambda_4=0$ and $\lambda_1$, $\lambda_2$ ,$\lambda_3$ can be found in the Jupyter notebook EVOLUTIONQED.ipynb accompanying this manuscript. In this notebook, the eigenvalues are calculated as the three nonzero roots of the characteristic polynomial of the kernel matrix $\matriz{P}^{(0,1)}$. 

\subsection{Combining QED and QCD evolution}
\label{qedyqcd}
Equipped with an operator for the evolution of QED and another for the evolution of QCD that satisfies two separate differential equations, we can attempt a combination of both effects. 
For simplicity, we will keep the following discussion only up to the LO order in QCD \footnote{Therefore, an extra $+{\cal{O}}(\as^2)$ is implicit  in all the expressions below.}, but the result can be expanded to NLO without difficulties. We start by proposing the following solution for the Eq.(\ref{eq:singlete}),
\beqn
\matriz{E}\left(Q, Q_0\right)=\matriz{E}^{QCD}\left(Q, Q_0\right)\matriz{E}^{QED}\left(Q, Q_0\right),
\label{eq:soluCE}
\eeqn
where the operators $\matriz{E}^{QCD}$ and $\matriz{E}^{QED}$ are solutions of the equations Eq.(\ref{eq:ecuacionSingleteQCD})(only the LO) and Eq.(\ref{eq:ecuacionSingleteQED}) respectively. Taking the derivative with respect to $t$, and using the evolution equations of the coupling constants Eq.(\ref{eq:copuling}), we get
\beqn
\frac{d\matriz{E}\left(Q, Q_0\right)}{dt}&=& \{(\al \matriz{P}^{(0,1)}+\as \matriz{P}^{(1,0)})\\ \nn
\quad &-&  \frac{1}{\beta_0\beta'_0} \left[ \ln{\frac{\as}{\aso}} \beta'_0\al\right][\matriz{P}^{(1,0)},\matriz{P}^{(0,1)}]\} \matriz{E}\left(Q, Q_0\right)+\mathcal{O}(\as^n\,\al^m)\,,
\eeqn
with $n+m=3,\, (n,m \neq 0)$.
We note that by neglecting the last term on the right-hand side we can get the solution we were looking for, since QED and QCD effects decouple. This term can be neglected since it is of order $\mathcal{O}(\al\,\as)$. As we mentioned in the section \ref{sec:solutions}, the kernel matrices of QED ($\matriz{P}^{(0,1)}$) and QCD ($\matriz{P}^{(1,0)}$) do not commute, therefore, by posing in Eq.(\ref{eq:soluCE}) the operators $\matriz{E}^{QCD}\,\matriz{E}^{QED}$ in different order we arrive at a different possible solution. However, by inverting the operators, the only change is a sign in the commutator term of order $\mathcal{O}(\al\,\as)$ that we drop. In fact, by developing the operators in power series of the couplings (using that, $\ln{\frac{\as}{\aso}}\simeq -\as \, \beta_0 \, \ln{\frac{Q^2}{Q^2_0}}$ and $\ln{\frac{\al}{\al_0}}\simeq -\al \, \beta'_0 \, \ln{\frac{Q^2}{Q^2_0}}$), it can  be shown that the solutions have the form \cite{Bertone:2013vaa},
\beqn
\matriz{E}^{QCD}\matriz{E}^{QED}= \sum (\matriz{A}\, \as +\matriz{B} \,\al)^n + \matriz{C} \, \al \, \as,\\
\matriz{E}^{QED}\matriz{E}^{QCD}= \sum (\matriz{A}\,\as +\matriz{B} \,\al)^n - \matriz{C} \, \al \, \as ,\nn
\eeqn 
where we keep only the lowest mixed order term, and we define
\beqn
\matriz{A}=\frac{1}{n!}\matriz{P}^{(1,0)}\, \ln{\frac{Q^2}{Q^2_0}} , \hspace{0.5cm} \matriz{B}=\frac{1}{n!}\matriz{P}^{(0,1)}\, \ln{\frac{Q^2}{Q^2_0}}, \hspace{0.5cm} \matriz{C}=\frac{1}{2}[\matriz{P}^{(1,0)},\matriz{P}^{(0,1)}]\,\mathrm{ln}^2\left(\frac{Q^2}{Q^2_0}\right).
\eeqn
Then, if we propose a symmetric solution of the form \cite{Bertone:2013vaa} 
\beqn
\matriz{\tilde{E}}=\frac{\matriz{E}^{QED}\matriz{E}^{QCD}+\matriz{E}^{QCD}\matriz{E}^{QED}}{2},
\label{Eq:Evolcombinado}
\eeqn 
we can eliminate the mixed order term, leaving us at a higher accuracy. If we calculate the derivative with respect to $t$ we get,
\beqn
\frac{d\matriz{\Tilde{E}}}{dt}&=& \{\al \matriz{P}^{(0,1)}+\as \matriz{P}^{(1,0)}\\ \nn
&-&\frac{1}{2\beta_0\beta'_0} \left[ \ln{\frac{\as}{\aso}}\beta'_0\al-\ln{\frac{\al}{\al_0}}\beta_0\as\right][\matriz{P}^{(1,0)},\matriz{P}^{(0,1)}]\}\, \matriz{\Tilde{E}}+\mathcal{O}(\as^n\,\al^m),
\eeqn
with $n+m=3,\, (n,m \neq 0)$.
Developing the logarithms in the power series of the couplings, we can see that the mixed order term cancels out, and we obtain the required solution,
\beqn
\frac{d\matriz{\Tilde{E}}}{dt}&=& (\al \matriz{P}^{(0,1)}+\as \matriz{P}^{(1,0)}) \matriz{\Tilde{E}}+\mathcal{O}(\as^n\,\al^m)\, .
\eeqn

\section{Phenomenological results}
\label{sec:result}
In this section, we will apply the evolution operators to parton distributions and study their phenomenological implications. Although we are not attempting to perform a global analysis, these results will be useful as a first visualization of the phenomenological effects of the QED corrections on the polarized parton distributions and the structure function. To carry out the analysis, we use as a probe for quarks and gluon pPDFs the DSSV18 set of helicity parton densities \cite{DeFlorian:2019xxt}. On the other hand, due to the lack of experimental polarized data, we do not count on a distribution for the polarized photon. Therefore, in order to analyze the impact of the QED corrections, we present different scenarios for the photon pPDF, based on the unpolarized distribution NNPDF3.1luxQED \cite{Bertone:2017bme}. In that work, they performed a global analysis complemented by the LUXqed constraint that relates the photon density distribution to the lepton-proton scattering structure functions \cite{Manohar:2016nzj,Manohar:2017eqh}. However, the NNPDF3.1luxQED fit is only valid for energies higher than $Q^2=2.72 \, \textup{GeV}^2$. To obtain the distribution at the initial scale we want ($Q_0^2=1 \, \textup{GeV}^2$), we simply evolve it to that scale. Having the photon PDF, we propose three possible schemes to model the polarized pPDF at an initial scale $Q^2_0$:
\begin{itemize}
\item \textbf{A} An extreme scenario, where we take the polarized pPDF identical to the unpolarized one $\Delta\gamma(x) = \gamma(x)$
\item \textbf{B} A moderate scenario, where we assume a linear polarization, $\Delta\gamma(x) = x \, \gamma(x)$ 
\item \textbf{C} The null scenario, where we take the photon pPDF to vanish at the initial scale.
\end{itemize}

Considering that the positivity constraint $|\Delta f(x)| < f(x)$ imposes a limit on the size of the polarized density distributions \cite{Soffer:1994ww}, scheme \textbf{A} proposes to take the photon pPDF (at the initial scale) as the upper limit of this constraint.
The scheme \textbf{B} assumes a simple linear relation between the polarized and unpolarized distributions, that is roughly reproduced by the quark densities. 
At last, in the scheme \textbf{C}, we study the case of setting $\Delta \gamma(x)$ to zero.
As stated before, we combine these three schemes with the NLO set of polarized distributions from \cite{DeFlorian:2019xxt} and observe how the distributions are affected in the evolution due to the QED corrections and the polarized photon density.

In order to analyze the results, we define the relative QED corrections as,
\beqn
\delta f=\frac{f_{withQED}-f_{noQED}}{f_{noQED}}
\label{eq:relativecorrec}.
\eeqn
We evolve the DSSV pPDFs supplemented with the photon pPDF from $Q_0^2=1 \, \textup{GeV}^2$ to $Q^2=800\textup{GeV}^2$. In the evolution, we use a variable flavor number scheme with a minimum of three flavors and set the charm and bottom threshold at $Q^2_{m_c}= 2 \, \textup{GeV}^2$ and $Q^2_{m_b}= 20.25 \, \textup{GeV}^2$ respectively (we do not take into account the top quark).

\begin{figure}[h]
	\centering
	\includegraphics[width=1\textwidth]{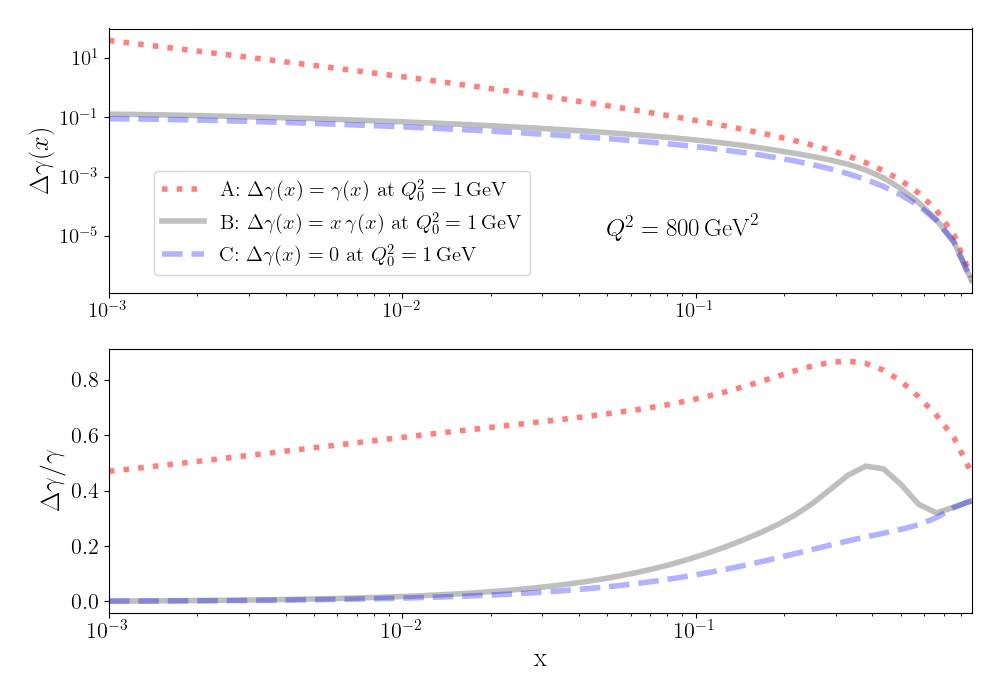}
	\caption{The photon pPDF (top) at $Q^2=800\, \textup{GeV}^2$. The ratio between polarized and unpolarized PDF (down). The colours indicate the different schemes chosen for the photon pPDF as explained in the main text.}
	\label{fig:photon}
\end{figure}

In figure \ref{fig:photon} (top) we show the photon pPDF evolved to $Q^2=800 \, \textup{GeV}^2$ for the different choices we made for the pPDF at $Q_0^2=1 \, \textup{GeV}^2$. As expected, the extreme scheme \textbf{A} differs significantly from the other scenarios, which provide similar results, since for them the photon distribution at high scales is mostly dominated by the emission from quarks more than from the assumption at the initial scale. In the bottom part of the figure \ref{fig:photon} we show the ratio between the polarized and unpolarized photon distributions. 

\begin{figure}[]
\centering
\begin{subfigure}{\includegraphics[width=80mm]{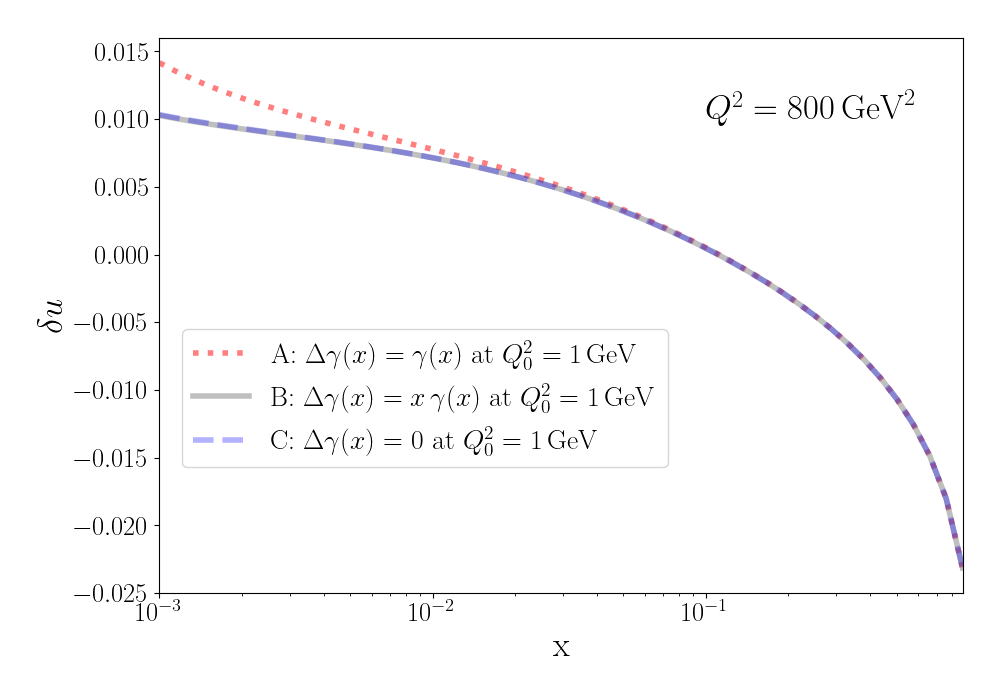}}\end{subfigure}
\begin{subfigure}{\includegraphics[width=80mm]{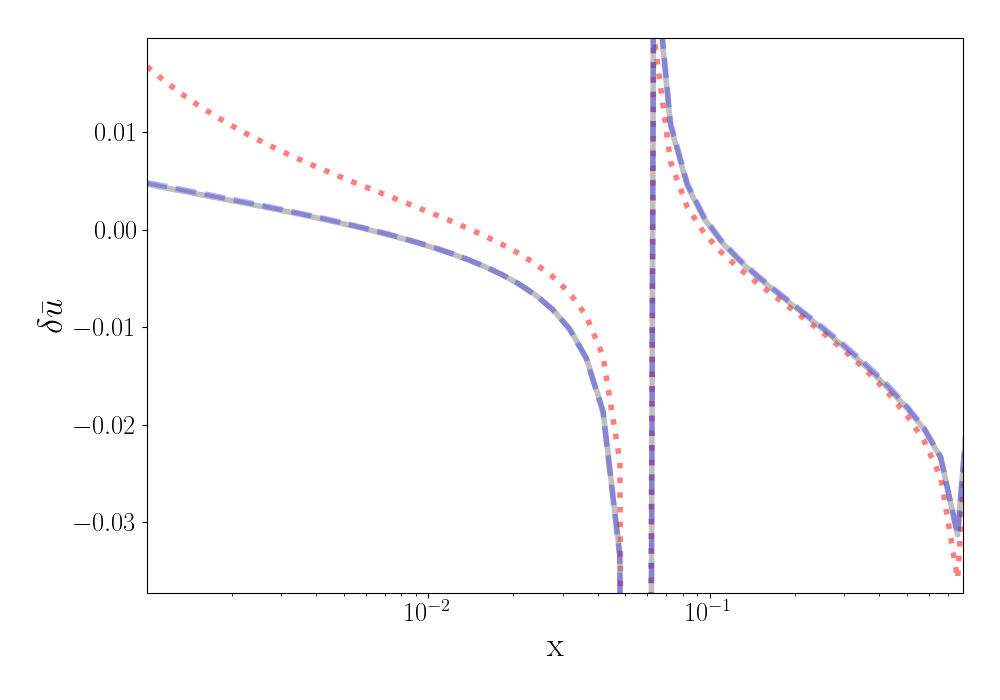}}\end{subfigure}\vspace{0mm}
\begin{subfigure}
{\includegraphics[width=80mm]{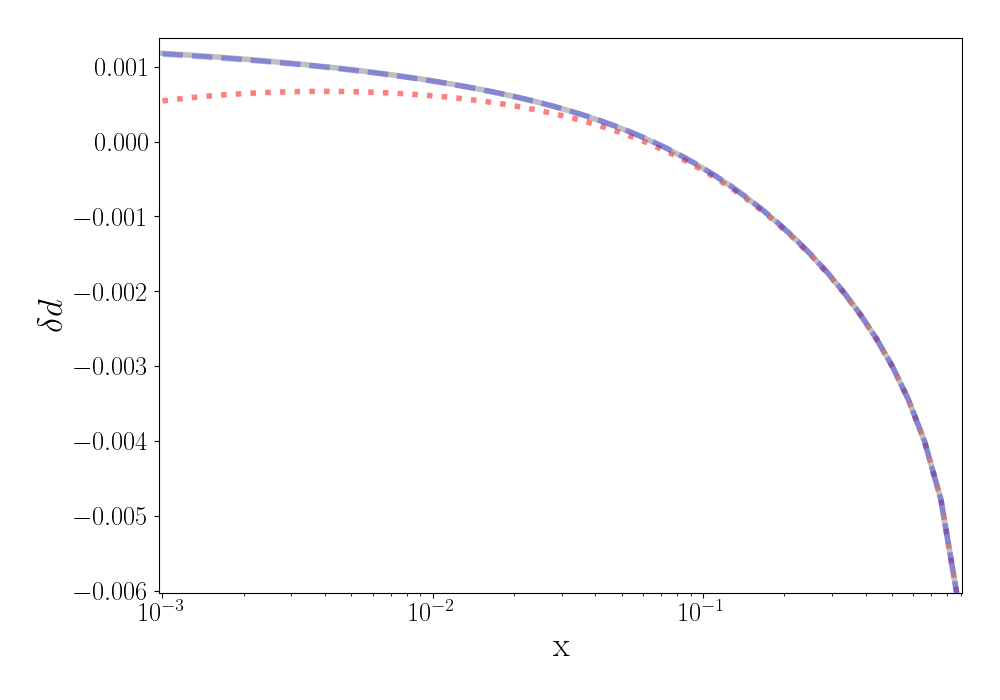}}\end{subfigure}
\begin{subfigure}{\includegraphics[width=80mm]{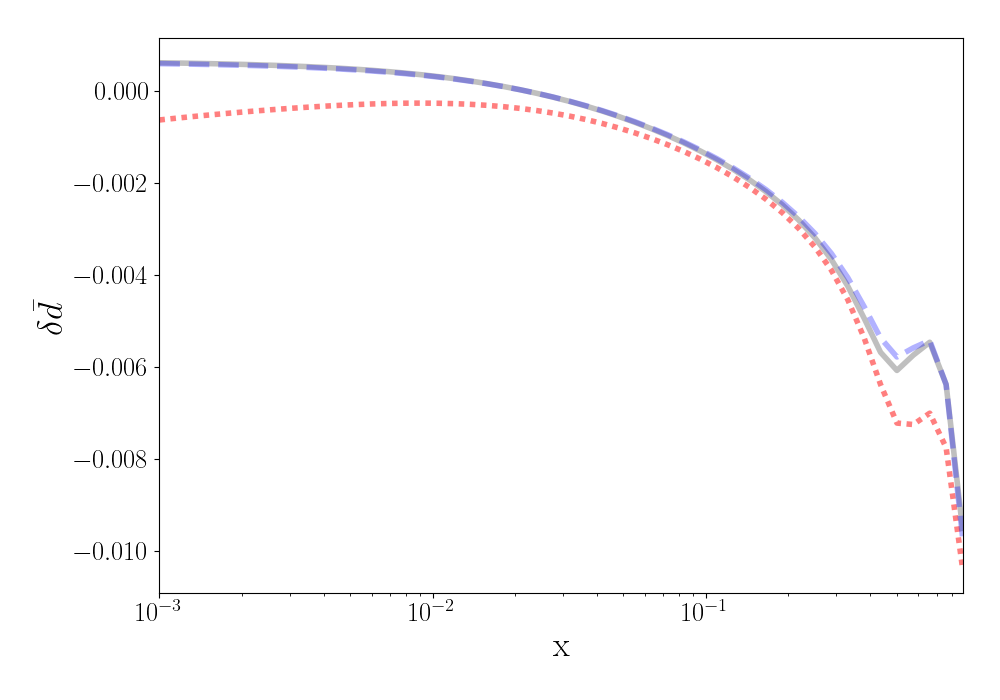}}\end{subfigure}
\begin{subfigure}{\includegraphics[width=80mm]{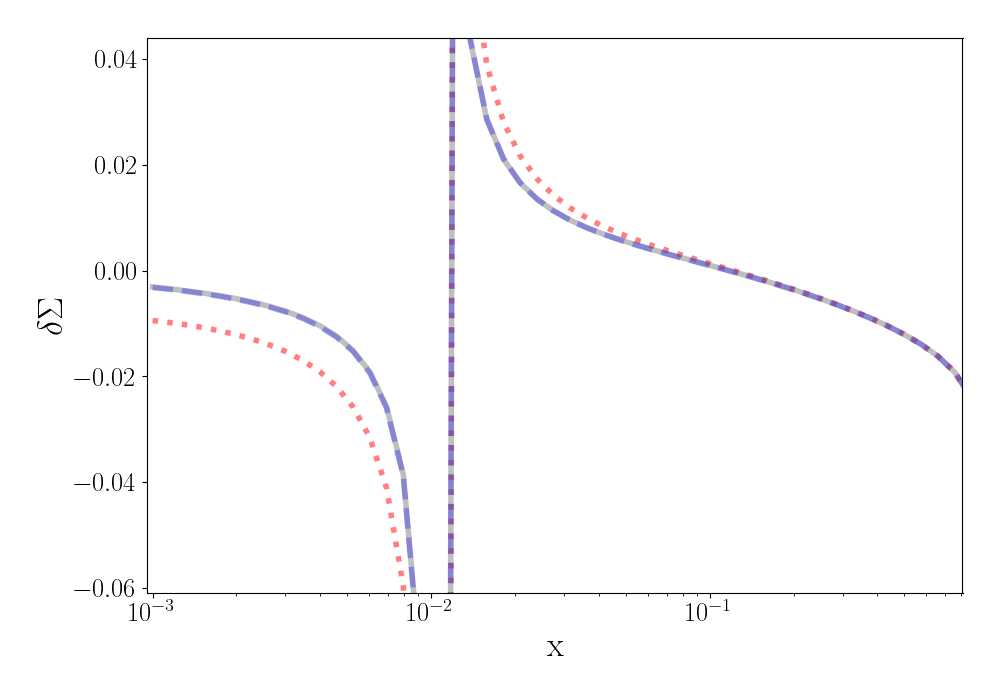}}\end{subfigure}
\begin{subfigure}{\includegraphics[width=80mm]{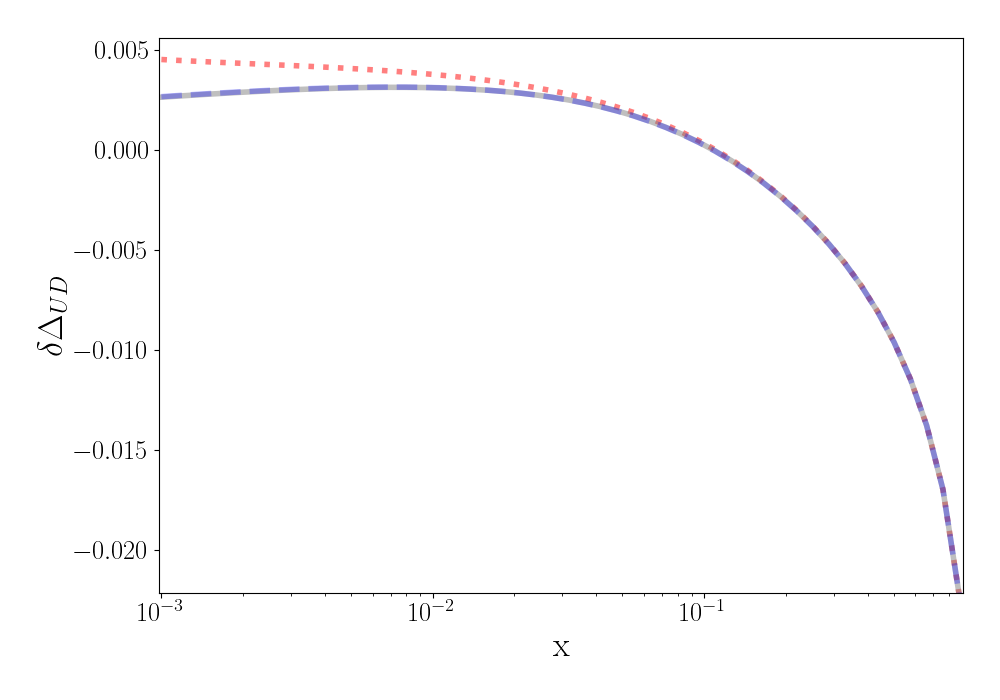}}\end{subfigure}
\begin{subfigure}{\includegraphics[width=80mm]{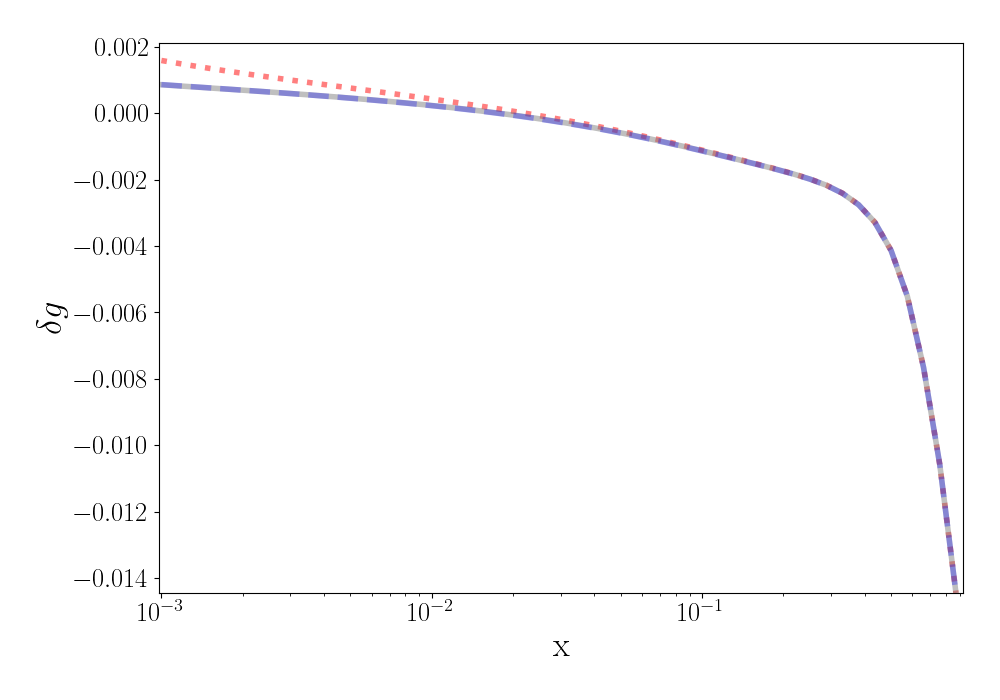}}\end{subfigure}
\caption{Relative QED corrections Eq.(\ref{eq:relativecorrec}) to the pPDFs at $Q^2=800\, \textup{GeV}^2$.} 	
\label{fig:partons}
\end{figure}
In the figure \ref{fig:partons} we show the relative corrections for the parton distributions of $u$-quark, $d$-quark, $\Sigma$, $\Delta_{UD}$ and gluon as a function of $x$. The QED corrections are large in the regions of $x$ where the partons distributions are small. The effect is typically of the order of the few percent, the same order of magnitude expected from the NNLO QCD corrections \cite{nachoborsa}.
In the cases of $\Sigma$ and $\bar{u}$ pPDFs, there is a change of the sign in the distribution, forcing the relative effect to be very large due to the smallness of the corresponding QCD distribution, Eq.(\ref{eq:relativecorrec}). 
We also show the results for the different scenarios chosen to model the photon pPDFs. In particular, the scheme \textbf{A} (red) produces significantly different results from the other schemes (scheme \textbf{B}: grey, scheme \textbf{C}: blue) in the small $x$ regions.
 
\begin{figure}[H]
	\centering
	\includegraphics[width=1\textwidth]{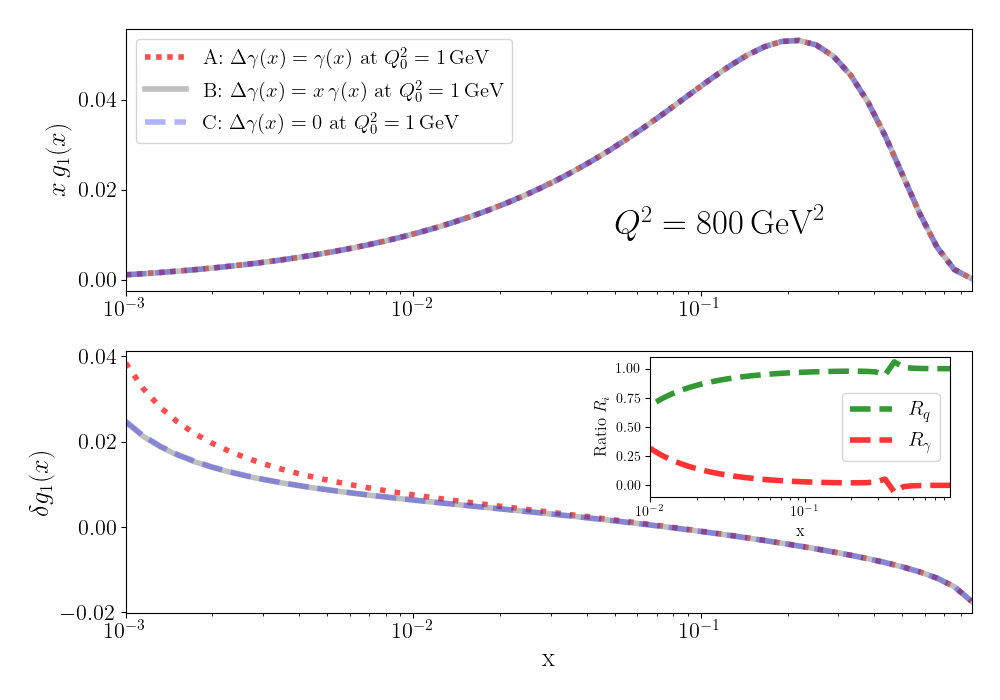}
	\caption{Structure function $x g_1$ (top) and its relative QED corrections (down) at $Q^2=800 \,\textup{GeV}^2$. On the inset, we show the contribution of each Wilson coefficient, $\Delta C_q^{(0,1)}$ and $\Delta C_\gamma^{(0,1)}$, represented by the ratios given by Eq.(\ref{eq:ratios}).}
	\label{fig:g1}
\end{figure}

In order to estimate the effect of QED corrections directly on physical observables, we first calculate the structure function $g_1$ to first order in QED. The function $g_1$ can be expressed in terms of the Wilson coefficients $\Delta C_i^{(i,j)}$ at different perturbative orders, in Bjorken $x$-space is expressed as,
\beqn
g_1=&\frac{1}{2}\sum_q e_q^2 \{(\Delta q+\Delta \bar{q})+ 2\,\as\,[\Delta C_q^{(1,0)} \otimes (\Delta q+\Delta \bar{q})+ \Delta C_g^{(1,0)} \otimes\Delta g] \label{Eq:gg11}
 \\
&+2\,\al\,\left[\Delta C_q^{(0,1)}\otimes(\Delta q+\Delta \bar{q})+\Delta C_\gamma^{(0,1)}\otimes \Delta \gamma \right]\}.\nn 
\eeqn 
Let us start by recalling the well-known coefficients to NLO in QCD \cite[and references therein]{Vogelsang:1996im,Lampe:1998eu}:
\beqn
\Delta C_q^{(1,0)}(x) &=&C_F \left[(1+x^2) \left[ \frac{\ln{1-x}}{1-x}\right]_+-\frac{3}{2}\frac{1}{(1-x)_+}-\frac{1+x^2}{1-x}\ln x+2+x-\left(\frac{9}{2}+\frac{\pi^2}{3}\right)\delta(1-x)\right], \nn\\ 
\Delta C_g^{(1,0)}(x) &=&2\,T_R  \, [(2x-1)\left(\ln{\frac{1-x}{x}}-1\right)+2(1-x)].\nn
\eeqn
To calculate the QED Wilson coefficients $\Delta C_i^{(0,1)}$, it is necessary to adjust the colour factors of the corresponding QCD coefficients, as described in Eq.(\ref{eq:colorajust}). Thus, we obtain
\beqn
\Delta C_q^{(0,1)}(x) &=&e_q^2 \left[(1+x^2) \left[ \frac{\ln{1-x}}{1-x}\right]_+-\frac{3}{2}\frac{1}{(1-x)_+}-\frac{1+x^2}{1-x}\ln x+2+x-\left(\frac{9}{2}+\frac{\pi^2}{3}\right)\delta(1-x)\right], \nn\\ 
\Delta C_\gamma^{(0,1)}(x) &=&2 \, N_C \, e_q^2 \, \left[(2x-1)\left(\ln{\frac{1-x}{x}}-1\right)+2(1-x)\right].\nn
\eeqn
In  figure \ref{fig:g1} we present the structure function $x g_1$ (with QED corrections) as a function of $x$ (top), and the relative QED corrections using Eq.(\ref{eq:relativecorrec}) (bottom). Again, the colours indicate the different schemes we chose to model the photon pPDF at the initial scale. Relative corrections are expected to be larger for values of $x$ where the function $g_1$ approaches zero. Nevertheless, we also observe corrections at the percent level in the region of $x$ where $ xg_1$ reaches its peak value.
On the other hand, in this particular region, there do not seem to be significant differences between the different schemes \textbf{A}, \textbf{B} and \textbf{C}. This suggests that the main correction in $g_1$ arises from the modification of the quark densities due to QED effects and, primarily, due to the incorporation of the coefficient $\Delta C_q^{(0,1)}$ in Eq.(\ref{Eq:gg11}).
In order to compare the size of the contributions of the Wilson QED coefficients, we defined the ratios,
\beqn
R_q=\frac{\Delta C_q^{(0,1)}\otimes(\Delta q+\Delta \bar{q})}{\Delta C_q^{(0,1)}\otimes(\Delta q+\Delta \bar{q})+\Delta C_\gamma^{(0,1)}\otimes \Delta \gamma} \,\hspace{0.5cm} \textup{ and } \, \hspace{0.5cm} R_{\gamma}=1-R_q. \hspace{1cm}
\label{eq:ratios}
\eeqn
In the bottom right-hand side of the figure \ref{fig:g1} we show the $R_i$ ratios as a function of $x$. Note that the contribution of $\Delta C_q^{(0,1)}$ is significantly higher than that of $\Delta C_{\gamma}^{(0,1)}$. 
The study presented in \cite{Borsa:2020lsz} estimates the potential impact of the EIC data on both the helicity proton distributions and the $g_1$ structure function. The results show that the expected experimental precision is comparable to the QED corrections obtained in the present work.
\section{Conclusions}
\label{sec:conclusions}
In this paper, we present the first-order polarized kernels in QED. Working in Mellin space, we solved the partonic evolution equations, including the photon pPDF, at ${\cal O}(\alpha )$ and ${\cal O}(\as^2)$. After introducing three possible scenarios for the unknown polarized photon distribution, we first analyze the phenomenological impact of the polarized parton distributions, using the set of pPDFs DSSV18 as a test probe. Finally, we calculate the QED corrections to the structure function $g_1$. We found that the main correction in $g_1$ arises from the modification of the quark densities due to QED contributions and from the incorporation of the $\Delta C_q^{(0,1)}$ coefficient in Eq.(\ref{Eq:gg11}). We show that the QED effects affect both parton distributions and the structure function at the percent level, the same order of magnitude expected for the NNLO corrections in QCD \cite{nachoborsa} and for the precision of future EIC measurements \cite{Borsa:2020lsz}.


\begin{acknowledgements}
We thank Iván Pedron for the discussions and useful communications. This work is partially supported by CONICET and ANPCyT.
\end{acknowledgements}

%

\end{document}